\documentclass[aps,prl,twocolumn,preprintnumbers,a4paper,superscriptaddress]{revtex4}

\usepackage{graphicx}

\begin{document}

\title{Current-voltage characteristics and the zero-resistance state
  in a 2DEG}

\author{F.S. Bergeret}
\affiliation{Theoretische Physik III, Ruhr-Universit\"{a}t Bochum,
  D-44780 Bochum, Germany}

\author{B. Huckestein}
\affiliation{Theoretische Physik III, Ruhr-Universit\"{a}t Bochum,
  D-44780 Bochum, Germany}

\author{A.F. Volkov}
\affiliation{Theoretische Physik III, Ruhr-Universit\"{a}t Bochum,
  D-44780 Bochum, Germany}
\affiliation{Institute of Radioengineering and Electronics of the
  Russian Academy of Sciences, Moscow 103907, Russia.}

\date{\today}

\begin{abstract}
  We study the current-voltage characteristics \emph{(CVC)} of
  two-dimensional electron gases (2DEG) with microwave induced
  negative conductance in a magnetic field. We show that due to the
  Hall effect, strictly speaking there is no distinction between N-
  and S-shaped \emph{CVC}s. Instead the observed \emph{CVC} depends on
  the experimental setup with, e.g., an N-shaped \emph{CVC} in a
  Corbino disc geometry corresponding to an S-shaped \emph{CVC} in a
  Hall bar geometry in a strong magnetic field. We argue that
  instabilities of homogeneous states in regions of negative
  differential conductivity lead to the observation of zero resistance
  in Hall bars and zero conductance in Corbino discs and we discuss
  the structure of current and electric field domains.
\end{abstract}

\pacs{73.43.-f,73.43.Qt,73.50.Fq}

\preprint{$Revision: 1.13 $}

\maketitle

\begin{figure}
  \begin{center}
    \includegraphics[bb=28 195 547 628,width=7cm]{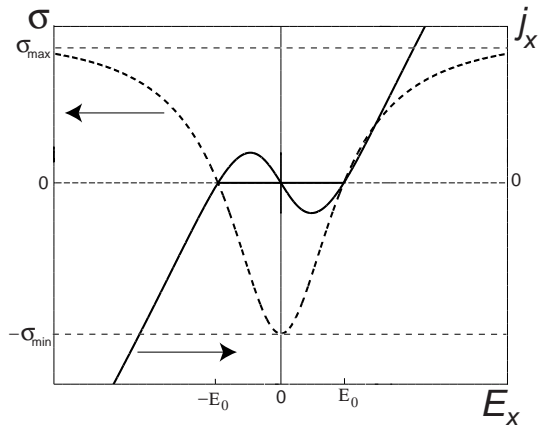}
    \caption{Assumed N-shaped CVC (solid line) and the
      corresponding dependence of the conductivity $\sigma$ on the
      electric field $E_x$ (dashed line). Both the current component
      $j_x$ and the conductivity $\sigma$ are zero at $E_x=\pm E_0$.
      The solid horizontal line ($j_x=0$, $|E_x|<E_0$) corresponds to
      a measurable part of the CVC in the Corbino disc measurements.
      \label{fig1}}
  \end{center}
\end{figure}

An interesting effect has been observed in recent papers \cite
{Klitzing,Zudov}: the resistance of a two-dimensional electron gas
(2DEG) subjected to microwave irradiation drops to zero in some
interval of an applied magnetic field $B.$ This zero resistance state
(ZRS) is achieved at low magnetic fields where at the measurement
temperatures Shubnikov-de Haas oscillations are weak.  Possible
mechanisms for this phenomenon are discussed in a number of papers
\cite{Philips,Yale,Andreev,Anderson,Shi,Raikh,Mikhailov}.
It was shown in Refs.~\cite{Yale,Shi,Raikh} that in the presence of
irradiation and a magnetic field the conductivity $\sigma _{xx}$ may
become negative in a weak electric field $E_{x}$.  In addition, using
a simple model, the authors of Ref.~\cite{Shi} have calculated the
current-voltage characteristic $I(V)$ of an irradiated system with a
density-of-states periodic in energy $\varepsilon $.  They have
demonstrated that not only regions with negative conductance $G$, but
also regions with positive $G$, may have negative differential
conductance $G_{d}=dI/dV$ on the current-voltage characteristic
\emph{(CVC)} curve. According to Ref.~\cite{Shi} the $I(V)$ curve is
N-shaped (or to be more exact it can be regarded as a chain of
the N-shaped \emph{CVC}s), similar to the one shown in
Fig.~\ref{fig1}.  This means that three voltages correspond to one
current. While the authors of Ref.~\cite{Shi} do not show in detail
how the ZRS arises in their approach, a phenomenological model for the
ZRS was proposed in Ref.~\cite{Andreev}. Here, the authors assumed
that the resistivity $\rho _{xx}$ depends on the current density $j$
in such a way that the dependence of electric field on current
density, $E_{x}=\rho _{xx}(j)j_{x}$, is N-shaped. This means that the
inverse dependence $j_{x}(E_{x})$ is S-shaped (three currents
correspond to one voltage, cf. Fig.~\ref{fig3}). It was shown that the
states corresponding to the part of the \emph{CVC} with negative
differential conductance are unstable. The authors assumed that a
stratification of the uniform current density occurs as a result of
this instability and two current domains with opposite current
directions arise in the Hall bar. With increasing total current
$I_{x}$ the relative width of the two domains changes and the electric
field $E_{x}$ remains zero, that is a ZRS is established in the
system.

We note that the results presented in
Refs.~\cite{Yale,Shi,Raikh,Andreev} are not completely new. Many years
ago the absolute negative conductance has been predicted in
Refs.~\cite{Ryzhii,Elesin,V'yurkov} (2 and 3DEG in a strong magnetic
field under irradiation) and in \cite{Epstein} (a superlattice under
irradiation). The instability of systems with N- or S-shaped
\emph{CVC} in the absence of a magnetic field and the possible types
of the electric field or current domains arising as a result of this
instability also has been studied three decades ago (see, for example,
the review \cite{KoganVolkovUs} and references therein). The
instability of a system with S-shaped \emph{CVC} in a magnetic field
was studied in Ref.~\cite{K}. As a result of the instability of a
homogeneous state, in systems with N-shaped \emph{CVC}, domains of
constant electric field arise whereas in systems with S-shaped
\emph{CVC} domains with constant current density appear.  In the
absence of a magnetic field it was shown that a wide domain of
constant electric field leads to a horizontal line on the \emph{CVC}
(zero differential conductance state) whereas a wide current domain in
the sample leads to the appearance of a vertical line on the
\emph{CVC} (zero differential resistance state)
\cite{KoganVolkovUs,KoganVolkovZh}.

Currently, there does not appear to be a consensus as to what kind of
\emph{CVC}, N- or S-shaped, will lead to the ZRS in the presence of a
magnetic field. In the present paper we address this issue and analyse
the form of the \emph{CVC} as well as the type of domains in different
experimental setups.  We consider a 2DEG in a magnetic field $B$ and
assume that the conductivity $\sigma _{xx}\equiv \sigma $ depends on
the electric field $E$ and the Hall conductivity $\sigma_{xy}\equiv
\sigma _{H}$ is independent of the electric field $E$.  We are not
going to analyse the origin of this type of non-linear behavior and
just note that this kind of dependence is discussed in
Ref.~\cite{Shi}. In a two-dimensional system, the electric field $E$
has two components $E_{x,y}$ so that the conductivity $\sigma $
depends on $E=(E_{x}^{2}+E_{y}^{2})^{1/2}$.  If one of two components
vanishes identically (for example $E_{y}=0$), as it happens in
measurements on the Corbino disc, then $j_{x}=\sigma (E_{x})E_{x}$. We
assume that the $\sigma (E_{x})$ dependence corresponds to an N-shaped
\emph{CVC} (see Fig.~\ref{fig1}). We will show that the form of the
\emph{CVC} in experiments on a Hall bar depends strongly on the
magnetic field, becoming S-shaped in the limit of strong Hall effect:
$\sigma _{H}\gg|\sigma|$.  This regime $\omega_c \tau_t\gg 1$, where
$\omega_c$ is the cyclotron frequency and $\tau_t$ is the momentum
relaxation time, is
relevant to the experiments under consideration. We note in passing
that in the experiments, due to the predominant forward scattering
nature of the disorder, the single particle lifetime $\tau_s$ is much
shorter than $\tau_t$ \cite{Klitzing}. In the following, we
will discuss the form of the \emph{CVC} and the structure of
non-homogeneous states (electric field and current domains) in
different experimental setups.

We consider a 2DEG in a magnetic field assuming first that the current
density $j$ and the electric field $E$ are homogeneous in the sample.
The current density $j_{x,y}$ depends on the electric field $E_{x,y}$
via the components $\sigma$ and $\sigma_H$ of the conductivity tensor,
\begin{equation}
j_{x,y}=\sigma (E)E_{x,y}\pm \sigma _{H}E_{y,x}.  \label{1}
\end{equation}
If the sample has the shape of a Corbino disc, the electric field
$E_{y}$ in the azimuthal direction is zero, and the \emph{CVC} shown in
Fig.~\ref{fig1} is N-shaped. If we consider a sample in the shape of a
Hall bar, then in general both components $E_{x,y}$ are finite.
However, the current component $j_{y}$ in the Hall bar is zero and
hence we can express the component $E_{y}$ through $E_{x}$:
$E_{y}=E_{x}\sigma _{H}/\sigma (E).$ Inserting it into the expression
for the current density $j_{x}$, we get the dependence of $j_{x}$ on
$E_{x}$,
\begin{equation}
j_{x}=\frac{\left(\sigma ^{2}(E)+\sigma _{H}^{2}\right)}{\sigma
  (E)}E_{x}.  \label{2}
\end{equation}
For the component $E_x$ of the electric field we get
\begin{equation}
|E_{x}|=\frac{|\sigma (E)|}{\sqrt{\sigma ^{2}(E)+\sigma _{H}^{2}}}E.
\label{3}
\end{equation}

\begin{figure}
  \begin{center}
    \includegraphics[bb=-224 -178 227 195,width=7cm]{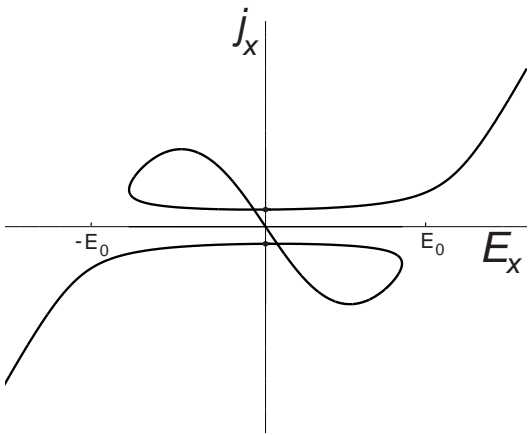}
    \caption{The CVC in the case of a weak magnetic field. Here
      $\sigma_{min}=\sigma_{max}=10\sigma_H$. The curve
        intersects the line $E_x=0$ at the points $j_x=0,\pm
        \sigma_H E_0$.\label{fig2}} 
  \end{center}
\end{figure}

\begin{figure}
  \begin{center}
    \includegraphics[bb=-224 -171 241 213,width=7cm]{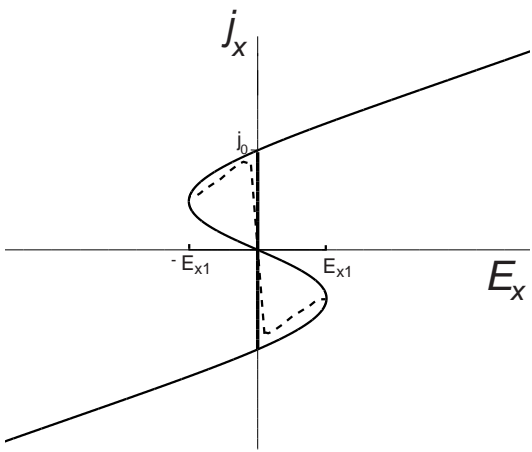}
    \caption{The S-shaped \emph{CVC} (solid line) for the case of large 
      magnetic field. Here
      $\sigma_{min}=\sigma_{max}=0.1\sigma_H$. The curve
        intersects the line $E_x=0$ at the points $j_x=0,\pm
        \sigma_H E_0$. The solid vertical line ($E_x=0$, $|j_x|<j_0$)
        corresponds to the stable measurable state on a Hall bar. For
        a discussion of the dashed line see text.\label{fig3}} 
  \end{center}
\end{figure}

Eqs.~(\ref{2}) and(\ref{3}) determine the form of the \emph{CVC} for
the Hall bar. We plot the resulting \emph{CVC} in Figs.~\ref{fig2} and
\ref{fig3} for the case of low and high magnetic field, respectively.
We see that in a weak field, $\sigma_{\mathrm{min,max}}\gg\sigma
_{H}$, the \emph{CVC} for almost all currents follows the form of the
$I(V)$ characteristic shown in Fig.~\ref{fig1}.  However at low currents,
$j_{x}\approx \sigma _{H}E_{0}$, the shape of the \emph{CVC} changes
drastically. This kind of the \emph{CVC} can not be assigned either to
the N- or S-shaped types of the \emph{CVC}. Associated with the loops is
a change of the Hall angle from small values at large negative $j_x$
to a value close to $\pi$ near $j_x=E_x=0$ and back to small values at
large $j_x$.

With increasing magnetic field the form of the \emph{CVC} changes and
finally, at $\sigma _{H}\gtrsim \sigma_{\mathrm{min,max}}$, is
transformed into an S-shaped $I(V)$ characteristic (Fig.~\ref{fig3}).
The $I(V)$ curve crosses the y-axis at the currents $j_{x}=0,$ $\pm
j_{0},$ where $j_{0}=\sigma_{H}E_{0}$.  The characteristic field
$E_{x1}$ at which $dj_{x}(E_{x})/dE_{x}=\infty $ is determined by
$E_{x}=E_{x1}(E_{1})$, where $E_{1}$ satisfies the equation
$E_{1}d\sigma (E_{1})/dE_{1}=-\sigma (E_{1})$ and is of the order of
$E_{x1}=aE_{0}\sigma_{\mathrm{min}}/\sigma_H$ ($a=2/3^{3/2}$ for a
parabolic dependence of $\sigma(E)$ on $E$ at small $E$). The
corresponding value of the current is $j_{x}(E_{x1})=j_0/\sqrt{3}$ in
the same approximation. In the limit of high magnetic field,
$\omega_c\tau_t\gg1$, the Hall angle does not change appreciably and
oscillates around $\pi/2$.

In order to check the applicability of this model to the experiments
\cite {Klitzing,Zudov}, it would be interesting to measure the
\emph{CVC} in two configurations. If the \emph{CVC} is measured on a
Corbino disc, the $I(V)$ curve should have the form shown in
Fig.~\ref{fig1}, that is the N-shaped form.  However due to
instability of a homogeneous state, a horizontal line ($j_{x}=0$)
should be observed in the interval $-E_{0}<E_{x}<E_{0}$ (zero
conductance state). In principle one can also observe the part with
negative absolute and differential conductance if short pulses of the
voltage are applied to the sample. This method of measuring the
\emph{CVC} of a homogeneous sample was used in studies of the Gunn
effect a long time ago (see the review \cite{KoganVolkovUs} and
references therein).  The duration of the voltage pulses should be
shorter than a characteristic time for the build-up of the electric
field domains, but longer than the period of the ac field applied to
the sample. In dc measurements the domains of the electric field
$E_{x}(x)$ should appear as it happens for example in the Gunn effect.
In the system under consideration the electric field has different
direction in each domain. A variation of the total voltage leads to a
change of the relative widths of the domains; the current remains
negligible (zero conductance state). Due to the Hall effect, the
electric field domains also carry current densities. Since the
strength of the electric field is given by the critical field $E_0$
where the Hall angle is $\pi/2$, the associated currents always flow
perpendicular to the direction of the electric field. Thus, in a
magnetic field, current and electric field domains exist
simultaneously.

If the \emph{CVC} is measured on a Hall bar, it has the form shown in
Fig.~\ref{fig3} with a vertical line in the interval
$-j_{0}<j_{x}<j_{0}$ (zero resistance state). The part of the $I(V)$
curve with negative conductance can also be measured if one applies
short current pulses.  The pulsed measurements would allow one to
determine a characteristic time required for the build-up of the
current filaments. It is worth noting that a switching of the system
into a non-homogeneous state has much in common with a first order
phase transition (see Ref.~\cite{KoganVolkovUs}). Therefore even in dc
measurements it is possible in principle to measure ``superheated''
and ``supercooled'' states which correspond to parts of the \emph{CVC}
with negative absolute, but positive differential conductance. As in
the case of a first order phase transition, the result of a particular
measurement will depend on factors that influence the nucleation of a
new phase (homogeneity of the sample, boundary conditions etc). If the
current filament is created in the middle of the sample, the
\emph{CVC} has the form shown in Fig.~\ref{fig3} by the dashed line
\cite {KoganVolkovUs}.  This means that the observed \emph{CVC} may
have a hysteresis: with decreasing current it goes along the upper
part of the \emph{CVC}, passes through the intersection point $j_{0}$
and reaches a point $j(E_{x1})$.  When $j_{x}$ decreases further, a
jump to a curve shown by the dashed line occurs and the \emph{CVC}
follows this line to the region of negative currents. In this
configuration, as it is noted in Ref.~\cite{Andreev}, the current
domains (current filaments) with opposite current directions are
elongated in the x-direction. The stratification of the current
density, as we mentioned above, is natural in systems with S-shaped
$I(V)$ characteristic. In the presence of a strong magnetic field the
electrical field domains arise simultaneously in the transverse
direction (the Hall field domains).  

In order to find the form of the electric field and current domains,
one needs to derive the full set of equations governing the space and
time dependence of all quantities of interest (electric field, current
density etc).  At present there is no such a theory. Here, we suggest
a simple phenomenological model that allows one to map this problem to
the problem considered earlier.  We assume that the Coulomb screening
length is less than the thickness of the 2DEG. In this case the relation
between the concentration of electrons $n$ and the electric field
$\mathbf{E}$ is given by the Poisson equation
\begin{equation}
\nabla\cdot\mathbf{E}=\frac{4\pi e}{\epsilon}\left(n-n_{0}\right),
\label{4}
\end{equation}
where $n_{0}$ is the electron density in the uniform case. Consider the
case of a Corbino disc. Then $E_{y}=0$ and the component $j_{x}$
equals
\begin{equation}
  j_{x}=\sigma (E_{x})E_{x}-eD\partial _{x}n+\frac{\epsilon}{4\pi}
  \partial_{t}E_{x}.  \label{5}
\end{equation}
Here the second term is the diffusion current and the third term is
the displacement current. This simplified model can not be applied
directly to the real experimental situation because it implies a local
approximation for the current density, that is, for example the
screening length $l_{scr}=\sqrt{D/(4\pi \sigma_{\mathrm{max}}
  /\epsilon )}$ is assumed 
to be longer than the mean free path (obviously this is not the case in
real samples). However this model in our opinion captures the main
ingredients of the system qualitatively. One can eliminate the
electron density from 
Eqs.~(\ref{4}) and (\ref{5}) and obtain for a stationary or steadily
moving domain of the electric field
\begin{equation}
  \frac{4\pi}{\epsilon}j_{x}=\frac{4\pi}{\epsilon}\sigma_{0}(E_{x})E_{x}
  - D\partial_{\xi \xi }^{2}E_{x} - \partial_{\xi}E_{x}
  \left(s-\frac{\sigma_{0}(E_{x})E_x}{en_{0}}\right).
\end{equation}
Here $\sigma _{0}(E_{x})=\sigma (E_{x})n_0/n$ and $\xi =x-st.$ We
assumed that $E_{x}(x,t)=E_{x}(\xi )$. One can see that Eq.~\ref{5} agrees
almost completely with an equation describing the Gunn effect (see
\cite{KoganVolkovUs}). The only difference is that the
$\sigma_{0}(E_{x})$ dependence is different in both cases; in
particular the symmetry point of the N-shape part of the \emph{CVC} in this
case is located at the origin of the coordinates ($E_{x},j_{x}$).
This means that the topology of solutions in the
($E_{x},\partial_{\xi}E_{x}$) plane remains unchanged and therefore
many conclusions about \emph{CVC}, forms of the electric field domains in
the Gunn effect are valid for our system (however domains in our
system are almost motionless). For instance, wide stationary domains
with the electric field equal to $\pm E_{0}$ are described by the
phase space trajectory (the separatrix)
\begin{equation}
  \frac{1}{r} \left[ {\cal E}_\zeta - \frac{1}{r} \ln \left( 1 + r
      {\cal E}_\zeta \right)\right] = U(1) - U(\zeta)\label{7}
\end{equation}
where $\zeta=\xi/l_{\mathrm{scr}}$, $r=E_0/(el_{\mathrm{src}}n_0)$,
${\cal E} = E_x/E_0$ is the normalized electric field,
$U({\cal E})=\int_{0}^{{\cal E}} d{\cal E}_1
j({\cal E}_1)/j_m$, $j_m = \sigma_{\mathrm{max}}E_0$, and
$j({\cal E}_1) = \sigma({\cal E}_1) {\cal E}_1$ is the
\emph{CVC}. The screening length $l_{\mathrm{scr}}$ in our model
determines the width of the domain wall. We will not study here in
detail the form of domains arising in the system. Note only that this
analysis can be carried out similarly to that presented in
Refs.~\cite{KoganVolkovUs,KoganVolkovZh}.

In summary, we have studied the form of the current-voltage
characteristic $I_{x}(V_{x})$ in the presence of non-linearity and
magnetic field. The Hall effect removes the distinction between N- and
S-shaped \emph{CVC}s characteristic for zero-field systems. Instead the
shape of the \emph{CVC} depends on boundary conditions imposed by the
experimental setup. We showed that an N-shaped \emph{CVC} in the Corbino disc
geometry corresponds for sufficiently strong magnetic fields to an
S-shaped \emph{CVC} in the Hall bar geometry. Homogeneous states with
negative differential conductivity are unstable to the formation of
domains. Again, due to the Hall effect these domains are characterised
by both a constant electric field and a constant current density. In
these domains the Hall angle is 90 degrees. As a result, we expect
the measurements of zero conductance in Corbino discs and zero
resistance in Hall bars to be manifestations of the same physical
phenomenon.

\textbf{Note added:} After the preparation of this manuscript we
became aware of Ref.~\cite{Yea03} that reports the observation of an
apparently zero-conductance state in a Corbino disc geometry.

\end{document}